\documentclass[biblatex]{lni}
\usepackage{amsmath}

\begin{document}
\title[]{A Methodology and System For Big-Thick
Data Collection}
\author[1]{Ivan Kayongo}{ivan.kayongo@unitn.it}{0009-0007-4429-7335}
\author[1]{Haonan Zhao}{haonan.zhao@unitn.it}{0000-0003-0825-7188}
\author[1]{Leonardo Malcotti}{leonardo.malcotti@studenti.unitn.it}{0009-0000-0589-6393}
\author[1]{Fausto Giunchiglia}{Fausto.giunchiglia@unitn.it}{0000-0002-5903-6150}%
\affil[1]{University\\of\\trento\\Italy}
\maketitle

\begin{abstract}
Pervasive sensors have become essential in research for gathering real-world data. However, current studies often focus solely on objective data, neglecting subjective human contributions. We introduce an approach and system for collecting big-thick data, combining extensive sensor data (big data) with qualitative human feedback (thick data). This fusion enables effective collaboration between humans and machines, allowing machine learning to benefit from human behavior and interpretations. Emphasizing data quality, our system incorporates continuous monitoring and adaptive learning mechanisms to optimize data collection timing and context, ensuring relevance, accuracy, and reliability. The system comprises three key components: a) a tool for collecting sensor data and user feedback, b) components for experiment planning and execution monitoring, and c) a machine-learning component that enhances human-machine interaction.
\end{abstract}










\vspace{-0.5cm}
\section{Introduction}
\vspace{-0.3cm}
Pervasive sensors are extensively utilised for data collection, which subsequently provides features that facilitate the understanding of the world \cite{yurur2014context}. This is often referred to as Big Data \cite{das2013big}. Although Big Data offers an objective perspective of reality, it is unable to elucidate the subjective motivations that drive an individual's actions. On the other hand is Thick Data, a category of data sources that are consistent with ethnographically aligned and carefully analysed observational data \cite{bornakke2018big}. 
Big data offers extensive quantitative insights, whereas thick data provides qualitative insights into human behaviour, experiences, and motives. When both are combined, to form Big-thick data, they provide a more comprehensive perspective of human needs and preferences to machines. One of the key elements of Big-Thick data is \textit{context} \cite{KD-2017-PERCOM,intille2003context,runyan2013smartphone}, which is  the situational setting of an individual that encompasses their internal condition within a common reference environment with others.

Current research on context data collection primarily focuses on context interruptions \cite{H-2017-Mishra}, capturing of user attention \cite{mehrotra2016my}, and enhancement of question response rates \cite{H-2021-Sun}. However, the current data quality does not meet the demands of big-thick data. A significant challenge arises from machines needing to pose a high volume of questions to humans, often resulting in low-quality responses \cite{H-2015-Mehrotra.a}. Certain methods also rely on fixed or random schedules that may not align with participants' availability or willingness to respond \cite{KD-2021-Zhang-putting}, thereby affecting the quality of the data.

In light of the above, our research objective in this paper is \textit{to develop a system that enables collection of high quality Big-Thick data, while ensuring minimal disturbance from the AI system}. Our aim is to collect quality Big-Thick data from both an objective and a subjective perspective, while monitoring this process. We  offer participants flexibility by incorporating a Machine Learning component which adaptively schedules context questions based on participants' availability and willingness to respond as a means to improve data quality. For monitoring, our system enables visualizing the collected data via a dashboard, thereby providing real-time feedback on the quality for example, missing sensor and question data. 

The system developed according to the methodology in this paper is an improvement and extension of iLog \cite{KD-2014-PERCOM}, which has been widely used in various data collection experiments and studies on human behavior. The first  experiment, SU2 (\textit{Smart University Two}), gathered data on students at the University of Trento, Italy. This data was used in \cite{KD-2021-Zhang-putting} to predict individuals’ behaviors, and in \cite{giunchiglia2018mobile}  to examine the impact of social media usage on students' academic performance. Another major data collection involved two experiments, DIV1 (\textit{Diversity One}) and DIV2 (\textit{Diversity Two}) \cite{giunchiglia2021worldwide},  which collected data on student behavior, mood, and food habits across eight countries. The extensions presented in this paper are motivated by the goal of having a fully integrated human-in-the-loop human machine interaction \cite{giunchiglia2022contextmodelpersonaldata, bontempelli2022lifelongpersonalcontextrecognition}.

The organization of this paper is as follows: Section \ref{relwork} provides an overview of the related work. Section \ref{Arc} presents our overall architecture. Section \ref{RepreContex} delineates the context model. Sections \ref{Monitoring} and \ref{planning} elucidate the monitoring and scheduling methods, respectively. Finally, Section \ref{con} offers the conclusion of the paper.

\vspace{-0.5cm}
\section{Related Work}
\label{relwork}

\vspace{-0.3cm}

The collection of big-thick data is mostly through; utilization of smartphone sensors and direct acquisition of information from participants. The former involves the use of sensors \cite{yurur2014context}, while the latter employs Time Use Diaries (TUDs) \cite{S-1939-Bowers}. In social sciences, the Experience Sampling Method (ESM) \cite{S-2014-Larson} is one of the principal methods for gathering information from participants to document their behaviors and this helps provide a ground truth as the data’s meaning is directly provided by the user. However it is beset with issues pertaining to the quality of responses \cite{KD-2023-Bison-answer,bison2024impacts}. A significant challenge resides in accurately determining the optimal timing for responses, a task complicated by the difficulties in observing respondent behavior when questions are scheduled at fixed intervals thus impacting its quality.

A number of platforms exist that collect and process big-thick data from smart devices. For instance, 
DemaWare2 \cite{stavropoulos2017demaware2} is a context-aware fusion system that employs OWL2 as its knowledge representation and reasoning language.
The contextual information provided is used to identify links between observations that signify the presence of complex activities.
Aware \cite{ferreira2015aware}, built on top of Beiwe \cite{onnela2021beiwe}, is an open-source mobile platform that generates user contexts from sensor data from smart devices and human digital questionnaires. 

To better understand the collected data, visualization can be employed. 
It effectively supports data exploration, insight communication, and data model improvement / understanding, for quality data. 
For example in \cite{samek2017explainable}, predictions of a deep learning model are explained through visualization
We leverage visualization not only to gain insights into the collected Big-Thick data but also to monitor the progress of the experiment for quality data collection.

\vspace{-0.6cm}
\section{Logical Architecture} 
\label{Arc}
\vspace{-0.3cm}
\begin{figure}[htp!]
\centering
\includegraphics[scale=0.6]{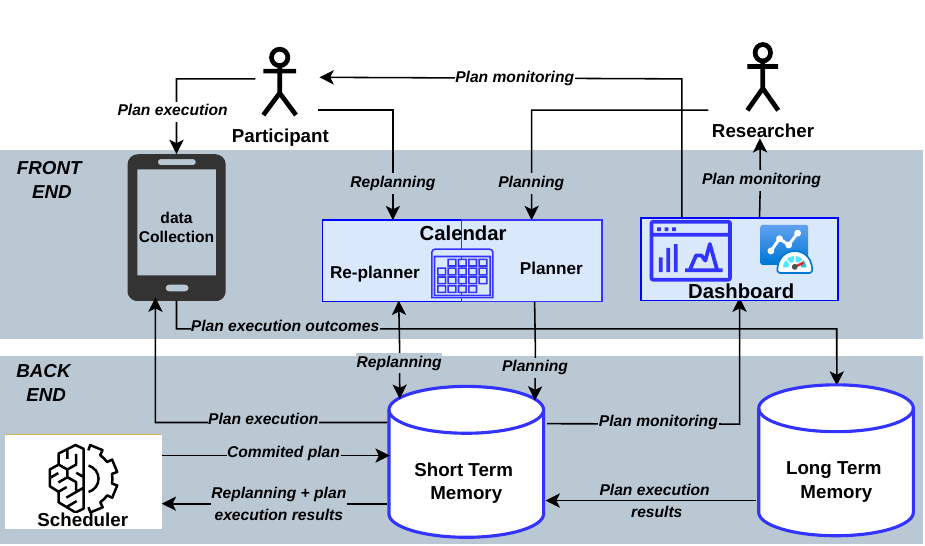}
\caption{ \centering Logical architecture}
\label{architecture}
\end{figure}
\vspace{-0.3cm}
\noindent
Figure \ref{architecture} depicts the overall system architecture; The light blue rectangles denote the primary components, the calendar and dashboard, with which individuals interact. Dark blue cylinders represent two databases, the first, termed the short term memory (STM) database, stores the plans from both researchers and participants, while the second, named Long Term Memory, the plan execution outcomes from participants. The orange rectangle represents the \textit{scheduler}, a machine learning module that modifies plans based on participants' data. The humanoid figures symbolize two roles, i.e., researchers and participants. A researcher is one seeking to gather context information from participants, while a participant is one that imparts their knowledge through the data collection app. The arrow lines illustrate the interactions between the different components, and the mobile phone serves as the tool for collecting data and enables the participant to receive context questions. 

The Big-Thick data collection process begins with the researcher’s planning. Using the Planner component within the Calendar module, the researcher schedules context questions and sensor data collection. This plan is stored in the STM database as scheduled actions, which are then sent to the data collection application at the designated times for participants to provide responses. Participants can modify the plan through the re-planner component in the Calendar module. Data collected during the execution of the plan is archived in the Long Term Memory database, while plan execution results are sent to the STM database. Furthermore, the scheduler component learns the optimal times to ask context questions based on the personal context and re-plan modifications from participants. It then sends these predicted times to the STM database to inform the planning process, ensuring the system adapts and responds effectively to changes.

Subsequently, the dashboard provides a means for both the researcher and participant to view and interact with the collected data. They can navigate through and filter the data, gaining valuable insights. For participants, this awareness of their lifestyle, as reflected in the data, can lead to positive behavioral improvements. Participants can also see how their data compares to that of other participants or to the limits set by the researcher.The dashboard provides researchers with a comprehensive overview of the experimental data, including any encountered problems. This enables them to promptly address these difficulties to ensure the seamless execution of the experiment for accurate data collection..

\vspace{-0.5cm}
\section{Representing Context} 
\label{RepreContex}
\begin{figure}[htp!]
\centering
\includegraphics[scale=0.45]
{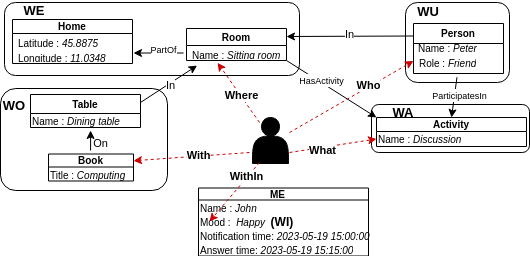}
\caption{\centering A motivating example of situational context model.}
\label{example}
\end{figure}

\vspace{-0.3cm}
The goal of asking context questions or collecting sensor data is to understand the personal context annotation from a participant. The annotated context used in this study is defined in \cite{KD-2017-PERCOM, KD-1993-giunchiglia}. As an example, let’s consider an afternoon at a student’s place where he is in a discussion with his friend. Figure \ref{example} shows this scenario as a knowledge graph, representing the personal context of the student. Each node represents an entity, e.g., person and room, with
their respective attributes with values; for instance, attributes of \textit{ME (whose context we are describing)} are $Class$, $Name$, $Mood$, \textit{Notification time} and \textit{Answer time} with the corresponding values as shown in Figure \ref{example}. Edges represent relations between entities, e.g., \textit{Sitting room} is $PartOf$ $Home$, whereas; \textit{Peter (person)} and the \textit{dining table (Table)} which are both $in$ the \textit{sitting room} which $HasActivity$ of a $discussion$ taking place.
The relation between context as a partial representation of the real world and the subject is represented in Figure \ref{example} as: 

  \makebox[330pt]{\textit{MyWorld = <Cxt, me >}}
  
where; \textit{Cxt} is the context (real world) of the subject, including elements in his local immediate surrounding and \textit{me} is the subject, represented as an entity with attributes and relations. 
The red arrows in figure \ref{example} show the relation between \textit{me} and \textit{Cxt} thus; 
\makebox[330pt]{\textit{ Cxt = < WA , WE , WI , WO , WU> }} \\
where:
\begin{itemize}
\item \textit{WA} is the temporal context from answering the question; “\textbf{W}h\textbf{A}t are you doing?”. In Figure \ref{example}, \textit{WA} is the main activity taking place, i.e. the \textit{discussion}.
\item \textit{WE} is the spatial context generated from the question “\textbf{W}h\textbf{E}re are you?”. In Figure \ref{example},  \textit{WE} shows the most relevant location, i.e. the \textit{sitting room}.
\item \textit{WI} is the internal context generated from the question "\textbf{W}hat mood are you \textbf{I}n?". As shown in Figure \ref{example}, it is the emotional state (mood) of \textit{me}.
\item \textit{WO} is the object context generated from the question “\textbf{W}hich \textbf{O}bject are you with?”. In Figure \ref{example}, \textit{WO} includes a few objects e.g., the \textit{dining table}, and the \textit{book} from which they are \textit{discussing}.
\item \textit{WU} is the social context generated from the question “\textbf{W}ho is with you (\textbf{U})?”. As shown in Figure \ref{example}, \textit{WU} focuses on \textit{me}’s friend \textit{Peter}.
\end{itemize}
The sensor data collected is used to add attributes to the context. For example in figure \ref{example}, the \textit{latitude} and \textit{longitude} attribute values in the \textit{Home} entity of the \textit{WE context} are sensor readings. Other collected sensor data includes; accelerometer, social media apps, Bluetooth devices among others, as detailed in \cite{giunchiglia2021worldwide}.

\vspace{-0.5cm}
\section{Monitoring the Plan Execution} 
\label{Monitoring}

\begin{figure}[htp!]
\centering
\includegraphics[scale=0.5]{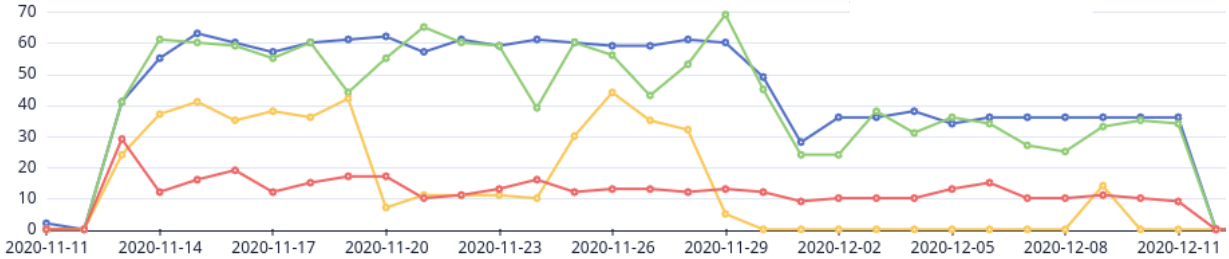}
\caption{ \centering Comparison of a participant's answers (green line) with three other participants (blue, red and orange lines)}
\label{compare_participant_answers}
\end{figure}

\vspace{-0.3cm}
The dashboard, as a visualization component offers flexibility to both the researcher and the participants in monitoring the execution of the experiment plan. It allows them to view and explore the collected big-thick data at any given time.
The dashboard helps them gain insights from the data as explained below.\\
\textbf{Researcher}: The researcher designs an experiment to determine which data to collect and when
and how to collect the data. Through the dashboard, he is able to view each participant’s data including the number of questions answered and sensor data collected per participant.
He can also compare the data collected between different participants and even filter it using options to get those with the least or most contributions. 
Not only does it facilitate him to gain insights in the collected data, but it also helps in the monitoring of the experiment progress to ensure that everything goes as planned. Details of the plan execution process including questions sent and data collected (answers and sensors) are displayed on graphs which in turn help in understanding if the plan is progressing as expected or if mitigation is needed, ensuring the collection of high-quality data.
\\
\textbf{Participant:} The participant registers to take part in an experiment and installs the data collection tool on a smart device. Through the dashboard, participants can navigate and explore their data to gain insights about themselves. For example, depending on the experiment's nature and questions, they can learn about their lifestyle, such as if they eat a lot of snacks or spend excessive time in a particular location. Such insights can highlight negative behaviors and prompt positive lifestyle changes. Participants can also compare their data against other participants to see how they fare in terms of data collected in the experiment. Figure \ref{compare_participant_answers} shows such an example, where one participant (green line) is compared to three others (orange, red and blue lines). The y-axis shows the number of questions answered whereas the x-axis shows the different experiment days. Such comparison prompts the participants improve on their contribution in the experiment.
\vspace{-0.2cm}
\\
The two functionalities explained above are some examples of the dashboard's functionalities; other key features include:
\begin{itemize}
    \item Alerting / notifying the researcher or the participant in case of inconsistencies in the collected data;
    \item Viewing participant response time and answer completion time and how these are affected by other factors for example social media, mood or location (sensor data);
    \item Enabling the participants to set personal goals and keep track of progress by using the collected Big-Thick data.
\end{itemize}

\vspace{-0.5cm}
\section{Scheduling} 
\label{planning}
\vspace{-0.3cm}
The scheduler is a machine learning component which learns and modifies the experiment plan by using a combination of the answers given by a participant and the modifications they make in the re-planner. Let's explain this with an example: Given that our participants are students, we try to refrain from sending questions while they are studying. To address this, we use the scheduler component to predict the periods when the participants are likely to be in class, thereby avoiding any disruption to their academic activities. 
Using data from the WeNet experiment described in \cite{giunchiglia2021worldwide}, we trained Random Forests, Decision Trees, Artificial Neural Networks, Logistic Regression, and Gaussian Naive Bayes on comprehensive training and testing sets encompassing all participants (170 students). We chose to binary encode study activities. Specifically, studying alone or with others and attending a classroom lecture are encoded as 1, and all other activities as 0. This  information was collected from time diaries questionnaire with the question: “What are you doing?”. The results are shown in Tab \ref{class} with the Random Forest classifier showing the highest prediction accuracy of 75.1\%. Other features that can be used for training include; social demographics (e.g gender, department), time (e.g hour, weekday), situational context (e.g location, interacting individuals), personality traits \cite{donnellan2006mini}, and mood.
\begin{table}
\centering
\caption{Prediction results of different machine learning classifiers on Wenet Data.}
\scalebox{0.67}
{
\begin{tabular}{|c|c c c c c c|}
    \hline
     \textbf{Classifier} & \textbf{Accuracy}    &\textbf{Kappa}&\textbf{Precision}&\textbf{Recall}&\textbf{F1 score}&\textbf{AUC}\\
    \hline
    Random Forest&0.7510&0.3832&0.6903&0.4491&0.5442&0.8146\\
    \hline
     Decision Tree&0.7234&0.3338&0.6091&0.4587&0.5233&0.6821\\
    \hline     Artificial Neural Networks&0.7209&0.3702&0.5781&0.5799&0.5790&0.7762\\
    \hline
     Logistic Regression&0.4557&0.3536&0.4892&0.0740&0.1286&0.6705\\
    \hline
     Gaussian Naive Bayes&0.6295&0.2388&0.4571&0.6366&0.5321&0.6676\\
    \hline
\end{tabular}
}
\label{class}
\end{table}

\vspace{-0.5cm}
\section{Conclusion}
\label{con}
\vspace{-0.3cm}
In this paper, we have presented a novel system for quality Big-Thick data collection, designed to enhance user experience and improve the quality and quantity of collected data. We have fostered a meaningful collaboration between humans and AI by granting participants the flexibility to choose their response times for context questions and sensor collections. Our system incorporates a visualization component, allowing both researchers and participants to monitor the experiment’s progress and gain insights from the data. Future work will focus on evaluating the system in real-world scenarios, comparing it with existing methods, and addressing potential ethical and privacy implications.


\vspace{-0.5cm}
\section*{Acknowledgment}
The research by Fausto, Ivan, and Leonardo were funded by the European Union's Horizon 2020 FET Proactive project “WeNet – The Internet of us”, grant agreement No 823783. The work by Haonan received funding from the China Scholarships Council (No.202107820038).


\printbibliography

\end{document}